\documentclass{aastex62}
\usepackage{graphicx}	
\usepackage{amsmath}	
\usepackage{amssymb}	
\usepackage{natbib}

\usepackage{float}
\usepackage[section]{placeins}

\graphicspath{{./}{figures/}}

\begin{document}

\title{Molecular Gas Heating and Modified Dust Properties in Active Galaxies: Growing Black Holes or Tidal Shocks?}

\author{Rebecca Minsley}
\affiliation{Department of Physics and Astronomy, Bates College, 44 Campus Avenue, Lewiston, ME 04240, USA}

\author{Andreea Petric}
\affiliation{Institute for Astronomy, 2680 Woodlawn Drive, Honolulu,HI, 96822,USA}
\affiliation{Canada-France-Hawaii Telescope, 65-1238 Mamalahoa Highway, Kamuela, HI, 96743, USA}

\author{Erini Lambrides}
\affiliation{Department of Physics \& Astronomy, Johns Hopkins University, Bloomberg Center, 3400 N. Charles St., Baltimore, MD 21218, USA}

\author{Aleksandar M. Diamond-Stanic}
\affiliation{Department of Physics and Astronomy, Bates College, 44 Campus Avenue, Lewiston, ME 04240, USA}

\author{Maya Merhi}
\affiliation{Lycoming College, 700 College Pl, Williamsport, PA 17701, USA}

\author{Marco Chiaberge}
\affiliation{Space Telescope Science Institute, 3700 San Martin Dr, Baltimore, MD 21218}

\author{Nicolas Flagey}
\affiliation{Canada-France-Hawaii Telescope, 65-1238 Mamalahoa Highway, Kamuela, HI, 96743, USA}

\begin{abstract}
We investigate if and how growing super-massive black holes (SMBH) known as Active Galactic Nuclei (AGN) and gravitational interactions affect the warm molecular gas and dust of galaxies. Our analysis focuses on the morphologies and warm ISM properties of 630 galaxies at z $<$ 0.1. We use {\it{grizy}} images from the Pan-STARRS survey to classify the galaxies into mergers, early mergers, and non-mergers. We use MIR spectroscopic measurements of emission from rotational H$_2$ transitions, dust and PAH features, and silicate emission or absorption lines at 9.7 $\mu$m to study how gravitational interactions impact the warm ISM in AGN and non-AGN hosts. 

We find that in AGN-hosts, the ISM is warmer, the ratios of H$_2$ to PAHs are larger, the PAH emission line ratios and silicate strengths have a wider range of values than in non-AGN hosts. We find some statistical differences between the $H_2$ emission of mergers and non-mergers, but those differences are less statistically significant than those between AGN and non-AGN hosts.

Our results do not establish a relation between the rate of BH growth and the warm ISM but point to highly statistically significant differences between AGN hosts and non-AGN hosts, differences that are not present with the same statistical significance between mergers and non-mergers. We speculate that the combination of triggering mechanisms, AGN orientations, and evolutionary stages that allow AGN to be classified as such in the MIR indicate that those AGN are energetically coupled on kpc scales to their host galaxies's warm ISM. Future optical and IR, spatially resolved spectroscopic studies are best suited to characterize this connection. 

\end{abstract}
\section{Introduction}
Growing super-massive black holes (SMBH) power compact, luminous, galactic centers known as Active Galactic Nuclei (AGN). At low redshifts, the most luminous AGN appear to be triggered by galaxy mergers \citep{ellis2019}. The changing gravitational potential of merging galaxies may allow a fraction of the gas to lose angular momentum and fall toward the center of the galaxy where it can feed any present SMBH or get compressed and induce star-formation \citep{hopkins2007,hopkins2016}. 
In this scenario some of the massive stars produced in the merger and the growing SMBH consume the surrounding ISM and inject energy into it through shocks, radiation, and turbulence. These processes increase the pressure in the gas, ionize part of it, and ultimately impede star-formation and SMBH growth \citep{hopkins2007,ellis2019}. Several parts of this scenario have been tested but the connections between AGN and their hosts' molecular gas on kpc scale are unclear. 

The ISM fuels the growth of stars and of SMBHs. The ISM is heated, ionized, and shocked by gravitational interactions, new stars, stellar deaths, and AGN winds, emission, and jets. Therefore, to understand a galaxy's evolution we need to study the amount, phase-structure, and temperature of its ISM. 

Spectroscopic observations in the MIR are a powerful way to identify and characterize AGN and a fraction of their ISM because in the MIR the AGN emission is (almost) not affected by extinction. Furthermore, the sources of MIR emission
(ionized interstellar gas, non-thermal radio sources, and dust particles) can be
empirically separated through a multitude of diagnostics \citep[e.g][]{genz1998, rig2002, armus2007, ds2010, petric2011}.
The warm (200-1000 K) molecular gas observed in the MIR makes up about 1\% to 30\% of the total molecular H$_2$ mass \citep{rous2007}. Studies of nearby radio galaxies (e.g. \citealt{ogle2007}) and mergers \citep{gui2009} showed that the warm molecular gas properties (mass, temperature, total luminosity relative to other coolants) can be used to estimate if and how the total ISM is affected by an AGN and/or by shocks associated with gravitational interactions. Important mechanisms responsible for exciting the MIR H$_{2}$ rotational transitions are: photoionization in star-forming regions \citep{gaut1976, bal1982}, AGN activity, X-ray heating, fluorescence induced by a non-thermal ultraviolet continuum, and shock heating by radio jets \citep{mor1988, lark1998, ogle2007, gui2012}. Rotational H$_{2}$ emission may also be powered by shocks in dense clumps of filaments associated with interacting galaxies \citep{app2006,ogle2007,gui2009}. Radio galaxies, some mergers, and some IR bright galaxies have enhanced H$_{2}$ emission with respect to the other MIR cooling lines (e.g. PAH, [Si II]) which suggests that shocks associated with supernova remnants, AGN, or tidal interactions can contribute significantly to the excitation of the H$_{2}$ rotational transitions \citep{hill2014, stir2014,petric2018, lamb2019}. 

Spatially resolved studies of radio jets environments in X-rays, radio, and molecular gas showed how jet-ISM interactions can propagate shocks into the ISM and affect star-formation \citep{lanz2015, ogle2014, app2018}.
In nearby Luminous Infrared Galaxies (LIRGS with L$_{\rm{IR}} > 10^{11}~\rm{L}_\odot$) about 20\% of warm H$_2$ heating may come from AGN, AGN hosts have warmer molecular gas than non-AGN hosts, and the sources with the biggest warm gas kinetic energies are mergers \citep{stir2014,petric2018}. The results of \citet{stir2014,petric2018} were based on relatively small numbers of AGN which made it difficult to compare AGN merger hosts to AGN non-merger hosts and SF merger hosts to SF non-merger hosts. 
More recently \citet{lamb2019} compiled $\sim$2000 MIR spectra taken by the Spitzer Space Telescope’s \citep{werner04} Infrared Spectrograph \citep[IRS,][]{houck2004} and found a strong statistical trend for AGN hosts to have warmer (by $\sim$200 K) warm molecular gas than non-AGN hosts, based on the H$_2$ S(5) - S(3) transitions. 
While it is clear that AGN host galaxies may have a warmer molecular gas component \citep{rig2002, ogle2007, zakamska2010, nesvadba2011, hill2014, stir2014, petric2018, lamb2019}, it is unclear what physical mechanism is responsible for heating up the molecular gas in most sources. Because the data we present lack both the spatial and spectral resolution to test the details of those physical mechanisms, in this paper, we limit our investigation to one question: can the higher temperatures measured by \citet{lamb2019} be associated with merging hosts and not driven by the presence of an AGN? To test this we study the morphologies of the $z < 0.1 $ galaxies in the \citet{lamb2019} and perform a complementary statistical analysis that includes non-detections. The answer to this is connected to a more physically interesting question: is it possible that the mechanisms which trigger AGN in mergers also heat up the gas, or do any AGN emissions, jets, winds, or outflows affect the temperature of the gas irrespective of the gravitational interactions of the host galaxies.



The remainder of this paper is organized as follows: in section 2 we present our sample of galaxies, the visual merger classification and the statistical tools we used to estimate the relative effect that mergers have on the temperature of the warm molecular ISM-component compared to AGNs. In section 3 we present our results for the warm molecular gas, the PAH size distribution and ionization, and the silicate strengths. In section 4 we discuss what our results may suggest about how galaxy interactions and growing SMBH impact the warm component of a galaxy's ISM. We adopt cosmological parameters of $h = 0.7$, $\Omega_{m} = 0.3$, $\Omega_{\Lambda} = 0.7$. 


\section{Sample and Analysis} \label{sec:style}
We classify a $z<0.1$ subsample of galaxies used in \citet{lamb2019} as mergers, early mergers, or non-mergers. We approximate the temperature of the warm molecular gas using the mid-infrared molecular emission lines of H$_2$ S(3) and the H$_2$ S(1). We use those temperatures together with the ratio of the H$_2$ S(3) to the 11.3 $\mu$m PAH to test if AGN and/or gravitational interactions affect the warm ISM. \citet{rous2007, zakamska2010, petric2018, lamb2019} find that the molecular emission lines ratios H$_2$ S(3) to H$_2$ S(1) are not affected by extinction.

\subsection{Targets}
The 2,015 galaxies analyzed by \citet{lamb2019} have redshifts between $z = 0.001$ and $z=4.27$ and were selected from the Spitzer archive\footnote{ $\rm{https://sha.ipac.caltech.edu/applications/Spitzer/SHA/}$} based on proposal topics related to active galaxies. The reduced spectra were provided by the Combined Atlas of Sources with Spitzer IRS Spectra \citep[CASSIS]{cas2011}, but further processed as described in \citet{lamb2019}. Our morphology analysis (see section 2.3) is based on visual classification using Pan-STARRS images, and therefore our sample only includes the galaxies at $z < 0.1$; given the typical spatial resolution of Pan-STARRS (0.7 - 0.8\arcsec) the visual classification becomes uncertain at higher redshifts \citep{nair&abr2010}. 

While \citet{lamb2019} found no obvious, systematic effects on the dust/gas analysis associated with the heterogeneous nature of the sample, a significant number of radio-loud AGN would skew our analysis because radio-loud AGN are mostly hosted by mergers \citep{chia2015}. We use the FIRST survey catalog \citep{vfirst} to compute rest-frame 1.4 GHz luminosities, using a spectral index of -0.7 as in \citet{lacy1993} and the radio-loud definition of \citet{ivezic2002} to determine that only 10 galaxies in our sample are radio-loud AGN. We did the analysis with and without the 10 RL galaxies and found that they had a minimal impact on the statistical tests. A future paper on the entire sample of \citet{lamb2019} will employ higher image quality optical data to expand the analysis to the higher redshift sources in the sample and will also benefit from on-going radio surveys with which to investigate the connection between radio selected AGN and the warm molecular gas. 

\subsection{MIR measurements}
Here we briefly present the MIR spectroscopic features used in our analysis. A more comprehensive description of those measurements is provided in \citet{lamb2019}. We first assume the temperature of the warm molecular gas is proportional to the ratio of two emission lines from rotational transitions of H$_2$:  S(3) (J=5-3 at 9.665$\mu$m) to S(1) (J=3-1 at 17.035 $\mu$m). We then use the 6.2, 7.7, and 11.3 $\mu$m PAH features to study the impact AGN and mergers may have on the sizes and ionization states of the dust grains. The relative strength of the 6.2 to the 7.7 PAH bands changes with grain-size distribution while the ratio of the 11.3 to the 7.7 PAH complexes are a function of their ionization  states \citep{lid2001,dli2007}. We use the strength of the 9.7 $\mu$m silicate feature as a proxy for dust obscuration. However, we note that differences in dust geometry relative to the sources of MIR emission may complicate the analysis \citep{zakamska2010}. 

We use the intensity of the 11.3 $\mu$m PAH and its ratio to the 24 $\mu$m flux as an estimate of the importance of star-formation in the studied galaxies. We acknowledge the debate around the validity of using 11.3 $\mu$m PAH as a robust tracer of SFR in AGN hosts \citep{smith07, ds2010, ds2012}.

The most direct methods employed to estimate the AGN contribution to the MIR emission in individual galaxies are the ratios of high to low ionization fine-structure emission lines. However, those lines require higher spectral resolution than that available to \citet{lamb2019}. Diagnostics from the low-resolution data are based on the dust properties (e.g. PAH features and dust continuum emission). For large samples of sources, diagnostics based on the PAH ratios or the MIR continuum slope match those from high to low ionization lines ratios \citep{petric2011}.

\begin{figure}[ht] 
\includegraphics[width=\linewidth]{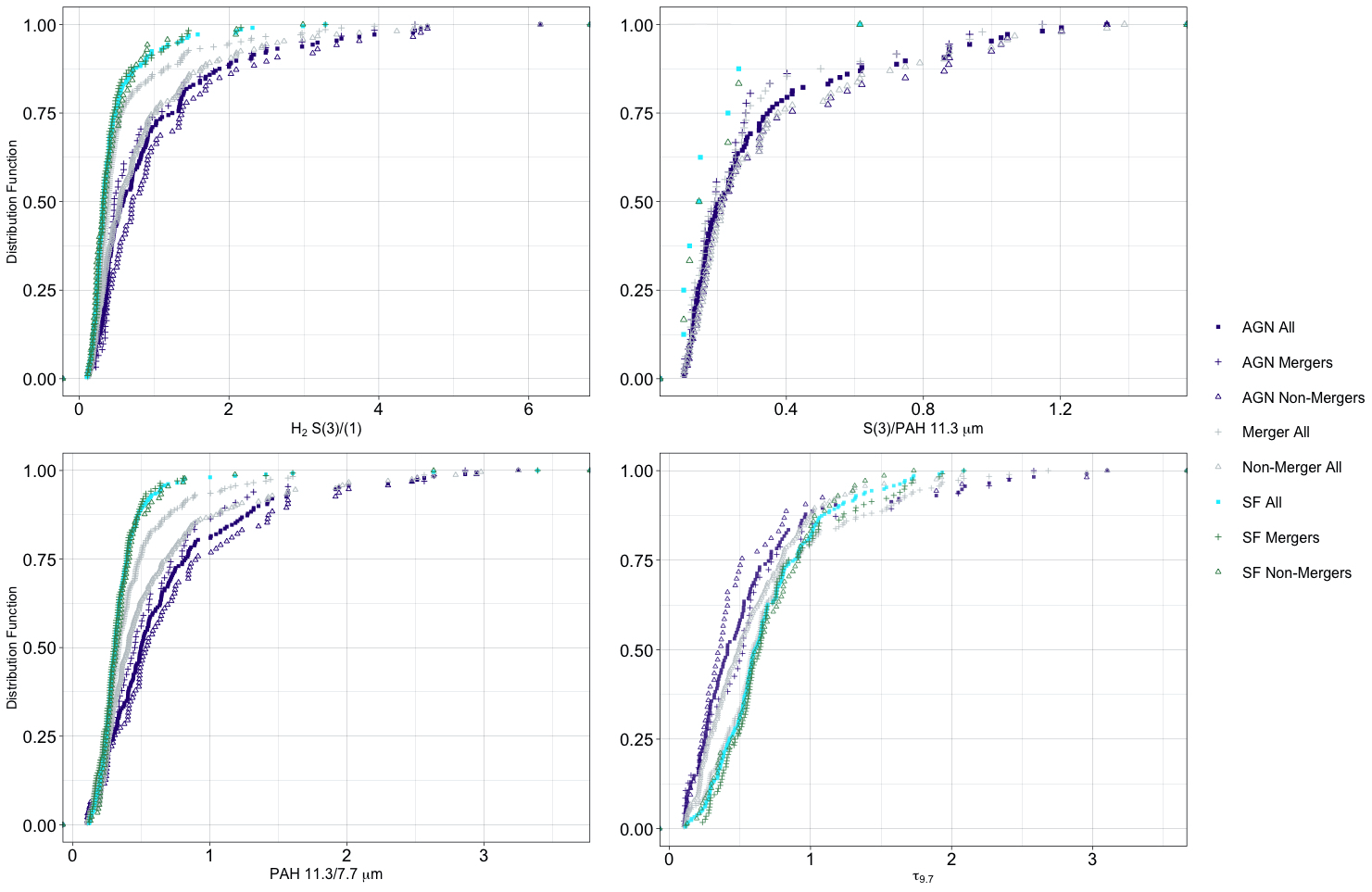}
\caption{Cumulative distribution functions for of H$_2$ S(3)/S(1) ratios (top left), H$_2$ S(3)/11.3 $\mu$m PAH ratios (top right), the 11.3 $\mu$m PAH / 7.7 $\mu$m PAH complex emission ratios (bottom left), and the silicate strengths $\tau 9.7\mu m$  (bottom right).  \label{fig:cdf}} 
\end{figure}

In figure \ref{fig:cdf} we present the cumulative distributions of the H$_2$ S(3) to H$_2$ S(1) ratios, H$_2$ S(3) to 11.3 $\mu$m PAH flux ratios, the 11.3 $\mu$m to 7.7 $\mu$m PAHs emission ratios, and the silicate strengths. 

The sample we present in this paper includes: 207 AGN (with 6.2 $\mu$m PAH EQW $<$ 0.27 $\mu$m), 114 AGN+SF composites (with 0.27 $<$ 6.2 $\mu$m PAH EQW $<$ 0.54 $\mu$m), and 264 SF only galaxies (with 6.2 $\mu$m PAH EQW $>$ 0.54 $\mu$m).


\subsection{Visual Classification}
 \begin{figure}[h!]
 \centering
 \includegraphics[width=0.9\linewidth]{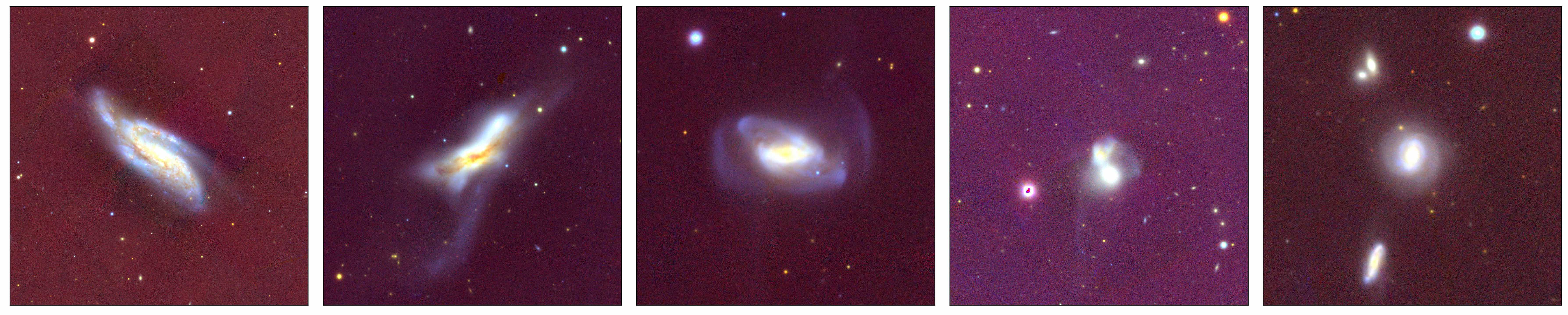}
 \caption{Pan-STARRs images of NGC 4088, NGC 0520, NGC 5218, NGC 4922 NED02, and UGC 07064 illustrating the different merger features we use for our classification: galaxy asymmetry, tidal tails, galactic shells, multiple nuclei and early/possible mergers for galaxies of similar brightness within 50 kpc of each-other.}
\label{fig:mergers}
\end{figure}

Our visual classification attempts to identify all galaxies in our sample that show signs of gravitational disturbances. We classify our targets as either non mergers, early mergers, or mergers. While our visual classification techniques was based on that of \citet{nair&abr2010} we also referred to \citet{larson2016,stir2013,bridge2010,zheng1997,petric2011} for their investigations of morphologies of nearby galaxies. 
We define non mergers as galaxies without any visible signs of a gravitation disturbance. We categorize early mergers as galaxies that have an apparent neighbor galaxy within $\sim$ 50 kpc radius of its its center. 

\citet{bb2020} point out the difficulties in identifying early mergers: tidal features are more evident in high stellar mass objects, with a high-gas fraction, and are also dependent on the geometry of the merger and the viewing geometry. Therefore while we present some of the possible statistical differences between early mergers, mergers, and non-mergers, we focus on statistical differences between mergers and non-mergers. We postpone a more focused discussion about early mergers to an investigation using multiple classification techniques. 
We use the term $merger$ to describe galaxies that show at least one of the following signatures of a gravitational disturbance: morphological asymmetries, tidal tails, shells, or multiple nuclei (see figure \ref{fig:mergers}). 

We were unable to classify 45 galaxies due to either defects in the stacked multi-filter images or uncertainty in the morphological features indicating either a disturbance due to gravitational interactions or an intricate dust geometry. Excluding these 45 galaxies, the total number of galaxies in our sample is 585: 248 mergers, 257 non mergers and 80 early mergers.
 These results are summarized in Table 1 along with median values of the aforementioned MIR measurements. 
 Three of the authors on this paper classified a subsample of 20 galaxies and were in excellent agreement: only two galaxies were assigned different classifications by one of the classifiers. 

\begin{deluxetable*}{cccc |c}[h!]
\centering
\tablecaption{Data Summary\label{tab:1}}
\tablecolumns{5}
\tablenum{1}
\tablewidth{0pt}
\tablehead{
\colhead{} & \colhead{Merger} & \colhead{Non-Merger} & \colhead{Early Merger} & \colhead{Total}
}
\startdata
\hline
& & Morphology classifications & &\\
\hline
AGN          & 73  & 104 & 30 & 207 \\
Star-forming & 139 & 91  & 34 & 264 \\
AGN/SF       & 36  & 62  & 16 & 114 \\ \hline
Total        & 248 & 257 & 80 & 585 \\
\hline
\hline
& & Medians of H$_2$ S(3) to S(1) ratios & &\\
\hline
AGN          & 0.5  & 0.7 & 0.3 & 0.6 \\
Star-forming & 0.3 & 0.3  & 0.3 & 0.3 \\
For all above & 0.4 & 0.5 & 0.3 & 0.4\\
\hline
\hline& & Medians of H$_2$ S(3) / 11.3 $\mu$m PAH ratios & &\\
\hline
AGN          & 0.10  & 0.11 & 0.13 & 0.11 \\
Star-forming & 0.01 & 0.01  & 0.01 & 0.01 \\
For all above & 0.02 & 0.05 & 0.03 & 0.02 \\
\hline
\hline
& & Medians of 11.3 $\mu$m PAH/ 7.7 $\mu$m ratios & &\\
\hline
AGN          & 0.4  & 0.5 & 0.4 & 0.5 \\
Star-forming & 0.3 & 0.3  & 0.3 & 0.3 \\
For all above & 0.3 & 0.4 & 0.3 & 0.3 \\
\hline
\hline
& & Medians of $\tau 9.7\mu$m & &\\
\hline
AGN          & 0.3 & 0.2 & 0.1 & 0.2 \\
Star-forming & 0.7 & 0.6  & 0.5 & 0.6 \\
For all above & 0.6 & 0.3 & 0.3 & 0.5 \\
\hline
\enddata
\end{deluxetable*}
\section{Results}
\subsection{Statistical Analysis}
We use the non-parametric, two sample tests, Logrank test and the Peto $\&$ Peto Generalized Wilcox test for censored data using the R statistical software package {\it{survival}} \citep{fin1985}, to investigate the relative role of mergers and AGN on the warm molecular gas and dust. Those tests look at differences between reconstructed distributions of values derived from measurements of detected features and upper limits and a maximum-likelihood formalism developed by \citet{km58}. This method was tested exhaustively by \citet{if1986} using Monte-Carlo simulations. 

In Table 2 we present the p-values from the Logrank, Peto, and Kolmogorov Smirnov (KS) tests which represent the likelihood that each of the two compared samples were drawn from a common population. The smaller the p-value the less alike the samples are. Values below 0.01 are significant, those between 0.01 and 0.1 are marginal and we only interpret them as possible hints for future analysis, values above 0.1 indicate no statistical differences between the samples compared, essentially telling us that the two populations are indistinguishable. 

\begin{deluxetable*}{lccccccc}
\rotate
\tablenum{2}
\tablecaption{A statistical table of warm ISM in active galaxies\label{tab:3}}
\tablecolumns{8}
\tabletypesize{\scriptsize}
\tablewidth{18pt}
\tablehead{
\colhead{Comparison} & 
\colhead{H$_2$ S(3) / S(1)} &
\colhead{H$_2$ S(3) / PAH $11.3~\mu$m } &
\colhead{\hspace{1pt} PAH $6.2 \mu$m\Big/PAH $7.7 \mu$m} & 
\colhead{\hspace{1pt} PAH $11.3 \mu$m\Big/PAH $7.7 \mu$m} & 
\colhead{\hspace{2pt} PAH $11.3 \mu$m} &
\colhead{\hspace{1pt} PAH $11.3 \mu$m$ \Big/f_\nu 24 \mu$m} &
\colhead{\hspace{1pt} $\tau_{9.7 \mu \rm{m}}$}
\\
\colhead{} & 
\colhead{\hspace{1pt} p$_{\rm{log}}$ \hspace{1pt} p$_{\rm{peto}}$\tablenotemark{a}} & 
\colhead{\hspace{1pt}p$_{\rm{log}}$ \hspace{1pt} p$_{\rm{peto}}$} &
\multicolumn{1}{c}{\hspace{4pt} p$_{\rm{log}}$ \hspace{1pt} p$_{\rm{peto}}$} & 
\multicolumn{1}{c}{\hspace{4pt} p$_{\rm{log}}$ \hspace{1pt} p$_{\rm{peto}}$} & 
\multicolumn{1}{c}{ \hspace{3pt} p$_{\rm{log}}$ \hspace{1pt} p$_{\rm{peto}}$ } &
\multicolumn{1}{c}{ \hspace{1pt}  p$_{\rm{ks}}$ \hspace{1pt} p$_{\rm{wilcox}}$}
}

\startdata
Mergers vs. Non Mergers & 
2e-4  \hspace{1pt} 8e-5  & 
7e-5 \hspace{1pt} 2e-5  &
0.3 \hspace{1pt} 0.3 & 
4e-4  \hspace{1pt} 5e-4  & 
0.1   \hspace{1pt} 0.1   & 
0.02 \hspace{1pt} 0.04 & 
5e-06 \hspace{1pt} 3e-06
 \\
&
(211 vs. 220) &
(246 vs. 256) &
(245 vs. 254) &
(246 vs. 256) & 
(246 vs. 256) & 
(211 vs. 219) &
(214 vs. 230)
\\
AGN vs. SF & 
4e-08 \hspace{1pt}  4e-09    & 
$<$2e-16 \hspace{1pt}  $<$2e-16 & 
2e-04    \hspace{1pt}  2e-04    & 
$<$2e-16 \hspace{1pt}  $<$2e-16 & 
0.04     \hspace{1pt}  0.04     & 
3e-08    \hspace{1pt}  3e-08    &
$<$2e-16 \hspace{1pt}  $<$2e-16 
\\
&
\hspace{2pt} (181 vs. 224) &
\hspace{2pt} (207 vs. 264) &
\hspace{2pt} (207 vs. 264) &
\hspace{2pt} (207 vs. 264) & 
\hspace{2pt} (207 vs. 264) & 
\hspace{2pt} (180 vs. 224) &
\hspace{2pt} (185 vs. 230)
\\
\\
Mergers  vs. Early Mergers  & 
0.7 \hspace{1pt} 0.7   &
0.7 \hspace{1pt} 0.6 &
0.09   \hspace{1pt} 0.1   & 
0.007 \hspace{1pt} 0.008 &
0.6   \hspace{1pt} 0.6   & 
8e-04 \hspace{1pt} 0.002 & 
0.004 \hspace{1pt} 5e-04
\\
&
\hspace{2pt} (211 vs. 71) &
\hspace{2pt} (246 vs. 80) &
\hspace{2pt} (245 vs. 78) &
\hspace{2pt} (246 vs. 80) & 
\hspace{2pt} (246 vs. 80) & 
\hspace{2pt} (211 vs. 71) &
\hspace{2pt} (214 vs. 69)
\\
Non Mergers  vs. Early Mergers  & 
0.04 \hspace{1pt} 0.03  &
0.03 \hspace{1pt} 0.02  &
1.0   \hspace{1pt} 1.0  &
0.9   \hspace{1pt} 0.9  &
0.4   \hspace{1pt} 0.4  & 
0.05  \hspace{1pt} 0.09 &
0.9   \hspace{1pt} 0.8 
\\
&
\hspace{2pt} (220 vs. 71) &
\hspace{2pt} (254 vs. 78) &
\hspace{2pt} (256 vs. 80) & 
\hspace{2pt} (256 vs. 80) & 
\hspace{2pt} (219 vs. 71) &
\hspace{2pt} (230 vs. 69)
\\
Mergers -- AGN vs. SF         & 
0.02     \hspace{1pt} 0.01    &
6e-08    \hspace{1pt} 2e-08   &
0.05     \hspace{1pt} 0.05    & 
2e-12    \hspace{1pt} 2e-12   & 
0.4      \hspace{1pt} 0.4     & 
3e-05    \hspace{1pt} 3e-05   &
5e-06    \hspace{1pt} 7e-06
\\
&
\hspace{2pt} (63 vs. 120) &
\hspace{2pt} (73 vs. 139) &
\hspace{2pt} (73 vs. 139) &
\hspace{2pt} (73 vs. 139) & 
\hspace{2pt} (73 vs. 139) & 
\hspace{2pt} (63 vs. 120) &
\hspace{2pt} (64 vs. 121)
\\
Non Mergers -- AGN vs SF   & 
5e-06 \hspace{1pt} 2e-06   &
8e-09  \hspace{1pt} 7e-09  &
0.01   \hspace{1pt} 0.01   & 
1e-08  \hspace{1pt} 9e-09  & 
0.1    \hspace{1pt} 0.1    & 
0.004  \hspace{1pt} 0.005  &
1e-10  \hspace{1pt} 4e-10 
\\
&
\hspace{2pt} (89 vs. 75) &
\hspace{2pt} (104 vs. 91) &
\hspace{2pt} (104 vs. 91) &
\hspace{2pt} (104 vs. 91) & 
\hspace{2pt} (104 vs. 91) & 
\hspace{2pt} (88 vs. 75) &
\hspace{2pt} (93 vs. 80)
\\
Early Mergers -- AGN vs SF  & 
0.1    \hspace{1pt} 0.1      &
0.01   \hspace{1pt} 0.008      &
0.1    \hspace{1pt} 0.1      & 
8e-06  \hspace{1pt} 8e-06    & 
8e-06  \hspace{1pt} 8e-06    & 
0.04   \hspace{1pt} 0.04     & 
0.007 \hspace{1pt} 0.02
\\
&
\hspace{2pt} (29 vs. 29) &
\hspace{2pt} (30 vs. 34) &
\hspace{2pt} (30 vs. 34) &
\hspace{2pt} (30 vs. 34) & 
\hspace{2pt} (30 vs. 34) & 
\hspace{2pt} (29 vs. 29) &
\hspace{2pt} (28 vs. 29)
\\
\\
AGN -- Mergers vs. Non-Mergers & 
0.3   \hspace{1pt} 0.3 &
0.07   \hspace{1pt} 0.06 &
0.3   \hspace{1pt} 0.3 & 
0.4   \hspace{1pt} 0.5 & 
0.6   \hspace{1pt} 0.5 &
0.9   \hspace{1pt} 0.7 &
0.005 \hspace{1pt} 0.01
\\
&
\hspace{2pt} (63 vs. 89) &
\hspace{2pt} (73 vs. 104) &
\hspace{2pt} (73 vs. 104) &
\hspace{2pt} (73 vs. 104) & 
\hspace{2pt} (73 vs. 104) & 
\hspace{2pt} (63 vs. 88) &
\hspace{2pt} (64 vs. 93)
\\
AGN  -- Mergers vs. Early-Mergers  & 
0.4 \hspace{1pt} 0.3 &
0.7 \hspace{1pt} 0.7 &
0.3 \hspace{1pt} 0.4 & 
0.3 \hspace{1pt} 0.4 & 
0.7 \hspace{1pt} 0.8 & 
0.1 \hspace{1pt} 0.2 &
0.4 \hspace{1pt} 0.2
\\
&
\hspace{2pt} (63 vs. 29) &
\hspace{2pt} (73 vs. 30) &
\hspace{2pt} (73 vs. 30) &
\hspace{2pt} (73 vs. 30) & 
\hspace{2pt} (73 vs. 30) & 
\hspace{2pt} (63 vs. 29) &
\hspace{2pt} (64 vs. 28)
\\
AGN -- Non-Mergers vs. Early-Mergers & 
0.02  \hspace{1pt} 0.01    &
0.1  \hspace{1pt} 0.1    &
0.9  \hspace{1pt} 0.9    & 
0.6  \hspace{1pt} 0.7    & 
0.6  \hspace{1pt} 0.6    & 
0.02 \hspace{1pt} 0.05   &
0.4  \hspace{1pt} 0.9
\\
&
\hspace{2pt} (89 vs. 29) &
\hspace{2pt} (104 vs. 30) &
\hspace{2pt} (104 vs. 30) &
\hspace{2pt} (104 vs. 30) & 
\hspace{2pt} (104 vs. 30) & 
\hspace{2pt} (88 vs. 29) &
\hspace{2pt} (93 vs. 28)
\\
\\
SF -- Mergers vs. Non-Mergers & 
0.9   \hspace{1pt} 0.9 &
0.9  \hspace{1pt} 0.9 &
0.7   \hspace{1pt} 0.8 & 
0.2   \hspace{1pt} 0.2 &
0.001 \hspace{1pt} 5e-04 & 
0.2   \hspace{1pt} 0.2 & 
0.2   \hspace{1pt} 0.2
\\
&
\hspace{2pt} (120 vs. 75) &
\hspace{2pt} (139 vs. 91) &
\hspace{2pt} (139 vs. 91) &
\hspace{2pt} (139 vs. 91) & 
\hspace{2pt} (139 vs. 91) & 
\hspace{2pt} (120 vs. 75) &
\hspace{2pt} (121 vs. 80)
\\
SF -- Mergers vs. Early Mergers & 
0.5   \hspace{1pt} 0.4   & 
0.5   \hspace{1pt} 0.5   &
0.9   \hspace{1pt} 0.5   & 
0.3   \hspace{1pt} 0.9   & 
0.1   \hspace{1pt} 0.004 & 
0.5   \hspace{1pt} 0.6   & 
0.004 \hspace{1pt} 0.007
\\
&
\hspace{2pt} (120 vs. 29) &
\hspace{2pt} (139 vs. 34) &
\hspace{2pt} (139 vs. 34) &
\hspace{2pt} (139 vs. 34) & 
\hspace{2pt} (139 vs. 34) & 
\hspace{2pt} (120 vs. 29) &
\hspace{2pt} (121 vs. 29)
\\
SF -- Non-Mergers vs. Early Mergers & 
0.5  \hspace{1pt} 0.5 &
0.5  \hspace{1pt} 0.5 &
0.8  \hspace{1pt} 0.5 & 
0.5  \hspace{1pt} 0.5 & 
0.7  \hspace{1pt} 0.5 & 
0.5  \hspace{1pt} 0.5 &
0.04 \hspace{1pt} 0.1 
\\
&
\hspace{2pt} (89 vs. 29) &
\hspace{2pt} (139 vs. 34) &
\hspace{2pt} (91 vs. 34) &
\hspace{2pt} (91 vs. 34) & 
\hspace{2pt} (91 vs. 34) & 
\hspace{2pt} (75 vs. 29) &
\hspace{2pt} (80 vs. 29)
\\ 
\\
AGN/SF -- Mergers vs. Non-Mergers &
\hspace{1pt} 0.02  \hspace{1pt} 0.01 &
\hspace{1pt} 0.01  \hspace{1pt} 0.02 &
\hspace{1pt} 0.08  \hspace{1pt} 0.1 & 
\hspace{1pt} 0.03  \hspace{1pt} 0.04 & 
\hspace{1pt}  0.4  \hspace{1pt} 0.4 & 
\hspace{1pt}  0.2 \hspace{1pt} 0.2 &
 \hspace{1pt} 0.3 \hspace{1pt} 0.2
\\
&
\hspace{2pt} (28 vs. 56) &
\hspace{2pt} (28 vs. 56) &
\hspace{2pt} (33 vs. 59) &
\hspace{2pt} (34 vs. 61) & 
\hspace{2pt} (34 vs. 61) & 
\hspace{2pt} (28 vs. 56) &
\hspace{2pt} (29 vs. 57)
\\
AGN/SF -- Mergers vs. Early Merger & 
\hspace{1pt}  0.01 \hspace{1pt} 0.009 &
\hspace{1pt}  0.09 \hspace{1pt} 0.1 &
\hspace{1pt}  0.04 \hspace{1pt} 0.06 & 
\hspace{1pt}  0.2  \hspace{1pt} 0.1 & 
\hspace{1pt}  0.8  \hspace{1pt} 0.7 & 
\hspace{1pt}  0.6  \hspace{1pt} 0.6 &
\hspace{1pt}  0.4  \hspace{1pt} 0.2
\\
&
\hspace{2pt} (28 vs. 13) &
\hspace{2pt} (28 vs. 13) &
\hspace{2pt} (33 vs. 14) &
\hspace{2pt} (34 vs. 16) & 
\hspace{2pt} (34 vs. 16) & 
\hspace{2pt} (28 vs. 13) &
\hspace{2pt} (29 vs. 12)
\\
AGN/SF -- Non vs. Early Merger & 
\hspace{1pt}  0.6 \hspace{1pt} 0.6 &
\hspace{1pt}  0.5 \hspace{1pt} 0.6 &
\hspace{1pt}  0.4 \hspace{1pt} 0.4 & 
\hspace{1pt}  0.6 \hspace{1pt} 0.7 & 
\hspace{1pt}  0.6 \hspace{1pt} 0.6 & 
\hspace{1pt}  0.8 \hspace{1pt} 0.8  &
\hspace{1pt}  0.8 \hspace{1pt} 0.9
\\
&
\hspace{2pt} (56 vs. 13) &
\hspace{2pt} (56 vs. 13) &
\hspace{2pt} (59 vs. 14) &
\hspace{2pt} (61 vs. 16) & 
\hspace{2pt} (61 vs. 16) & 
\hspace{2pt} (56 vs. 13) &
\hspace{2pt} (57 vs. 12)
\enddata
\tablenotetext{a}{The cited p-values from  the logrank, peto, and Kolmogorov Smirnov (KS) tests give the likelihood that the two samples were drawn from a common population. As such, the smaller the p-value the less alike the samples are. Values below 0.01 are highly significant, those between 0.01 but below 0.1 are marginal and we only interpret them as possible hints for future analysis, values above 0.1 indicate the samples are drawn from the same population.}
\end{deluxetable*}

\subsection{Effects of a growing SMBH on the warm molecular gas and dust}

We find that AGN hosts have higher H$_2$ S(3) to S(1) emission line ratios than non-AGN hosts suggesting higher molecular gas temperatures in AGN hosts. The median H$_2$ S(3) to H$_2$ (S1) ratio for all the AGN hosts is two times higher than that for SF hosts. We also find that AGN hosts have higher H$_2$ S(3) / 11.3 $\mu$m PAH ratios than galaxies without an AGN which suggests that the observed warmer may be related to the AGN emission. The largest and most statistically significant differences in terms of H$_2$ temperatures and H$_2$ S(3) / 11.3 $\mu$m PAH ratios are those between non-merger AGN hosts and non-merger SF hosts. 

The MIR emitting dust appears to also be affected by the presence of an AGN. We find statistically significant differences between the 11.3 $\mu$m to 7.7 $\mu$m PAHs emission ratios of AGN and non-AGN hosts with higher medians and a wider range of values for AGN hosts compared to non-AGN hosts. This result confirms, with a larger sample, the findings of \citet{smith07}. \citet{smith07} suggest that this may be due to differences in the ionization states of the PAHs: the 11.3 $\mu$m feature is thought to be produced by neutral PAHs, while the 7.7 $\mu$m feature arises primarily from PAH cations \citep[e.g][]{alla1999, dli2007}.  As discussed in \citet{smith07} harder radiation fields destroy PAH grains, therefore, if the grain or molecular carriers of the 7.7$\mu$m PAH feature are not the same as those of the 11.3 $\mu$m PAH feature they will by ionized and/or dissociated at different rates. The 7.7 $\mu$m PAH feature originates from smaller grains that may be destroyed easier than the bigger grains that emit the 11.3 $\mu$m PAH feature. 

We suggest the same explanation here although we think the location of silicates, PAHs, and H$_2$ relative to the AGN/SF regions may also play a role. 
As pointed out by \citet{zakamska2010, lamb2019} the ratio of the 11.3 to 7.7 $\mu$m PAH features is inversely proportional to the silicate strengths at 9.7 $\mu$m but not at all correlated with any of the H$_2$ properties (temperature, intensity, H$_2$/PAH). \citet{zakamska2010} suggest that the connection between the silicate strengths and the PAHs emission line ratios can be explained if the PAHs are located behind silicates and water ices in AGN dominated galaxies. A wider range of 11.3 to 7.7 $\mu$m PAH features ratios measured in AGN hosts may suggests that AGN hosts have more complex and diverse dust geometries than galaxies that do not host AGN.  

\begin{figure} 
\includegraphics[width=0.3\linewidth]{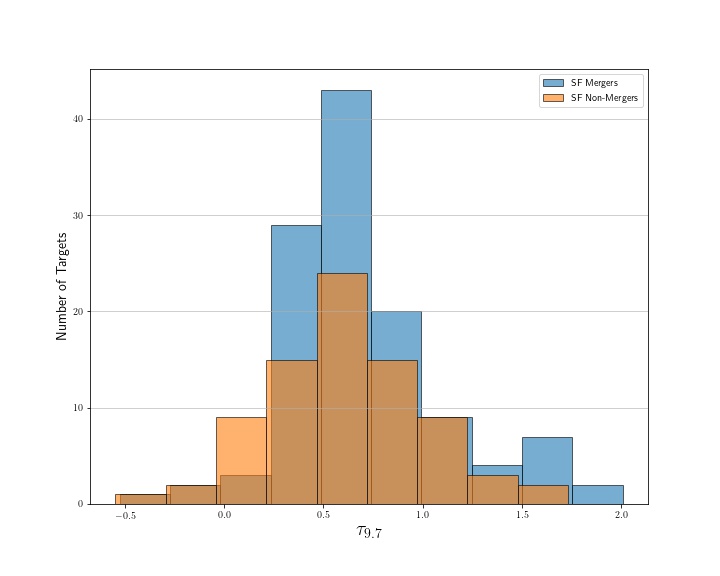}
\includegraphics[width=0.3\linewidth]{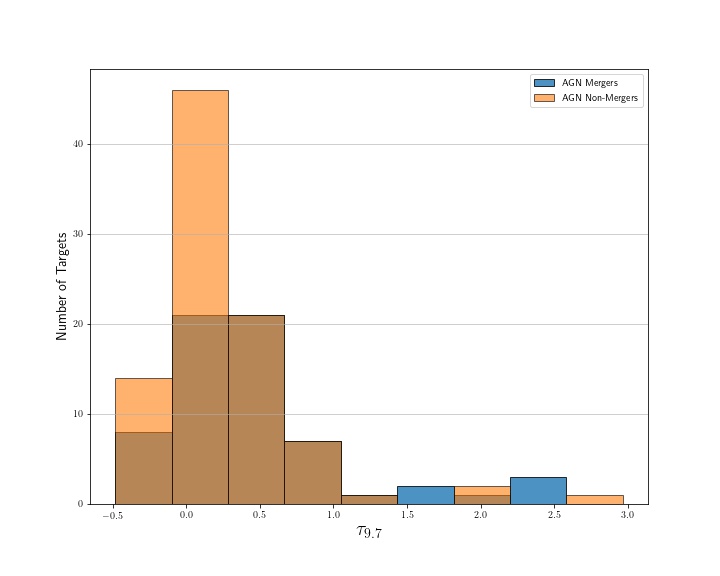}
\includegraphics[width=0.3\linewidth]{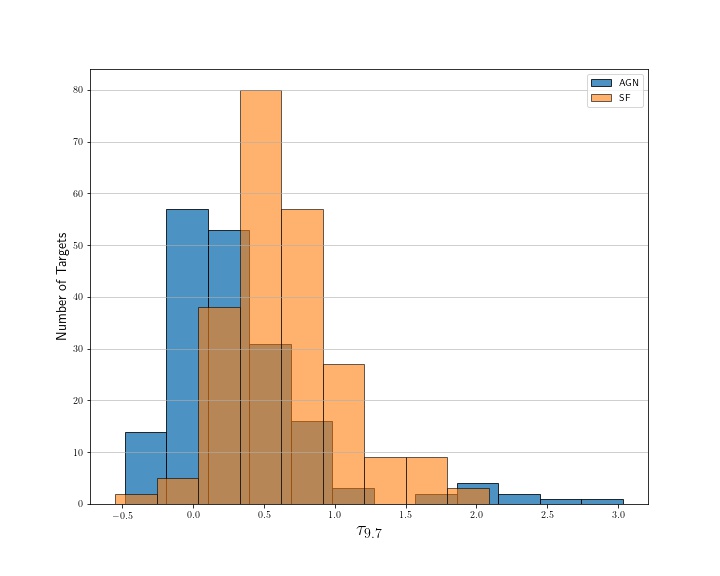}
\caption{Histograms of silicate strengths measured by \citet{lamb2019} for: (left) SF merger and SF non-merger galaxies, (center) AGN host-mergers and non-mergers, and (right) all the AGN and all the non-AGN hosts in our sample.  \label{fig:histoTau}} 
\end{figure}

In figure \ref{fig:histoTau} we compare the distributions of silicate strengths measurements in AGN and non-AGN hosts. We also compare the distributions of silicate strengths measurements in mergers and non-mergers. We find that the silicate strengths of non-AGN hosts differ significantly from those of AGN hosts. The median silicate strengths of AGN are about 3 times lower than those of non-AGN hosts and also span a wider range of values. This matches the finding of \citet{lamb2019} for a sample spanning a larger redshift range. The heterogenous nature of the targets in this study and the preponderance of (z$<1$) galaxies in our sample may suggests that we miss the most obscured AGN or that in the nearby universe (z$<1$) the fraction of highly obscured AGN is low \citep{Lacy2007}.

We also find statistically significant ($p < 2E-4$) differences between grain size distributions, as estimated by the L$_{6.2}$ $\mu$m / L$_{7.7}$ between AGN and non-AGN hosts.


\subsection{Effects of gravitational interactions on the warm molecular gas and dust}

We compare the warm molecular gas temperatures of all the mergers in our sample to those of all the non-mergers and find that their median H$_2$ S(3) to S(1) ratios are similar, but their distributions are different (the p-value for this comparison are 5E-6 and 2E-6). However, the differences in terms of the warm molecular gas temperatures among AGN only (AGN mergers vs AGN non-mergers) are not highly significant (p-values $\sim$ 0.03) and they are non-existent among non-AGN hosts (p~values $\sim$ 0.9). Among AGN+SF composites we do find slightly more statistically significant differences between mergers and non-mergers (p~values $\sim$ 0.01).

We also find that the distribution of PAH emission features flux ratios seems to differ significantly between mergers and non-mergers but the statistical strength of those differences decreases sharply when we compare sub-samples: AGN mergers vs AGN non-mergers or SF-mergers vs SF non-mergers. The mergers in our sample have deeper absorption features than non-mergers or early-mergers for both AGN and non-AGN hosts. In a future paper we will analyze the morphologies of the complete sample from \citet{lamb2019} to increase the number of galaxies available for analysis, and assess if/how stellar versus ISM mass may affect our findings. 

\section{Discussion}
In this work we analyze the morphologies and warm ISM properties, as estimated from $griz$ imaging with Pan-STARRS and MIR spectroscopic features, of 630 galaxies at z $<$ 0.1. We use non-parametric two sample tests for censored data as implemented in the R statistical software package $asurv$ to compare galaxies that host AGN to those that appear not to host AGN. We use the same tests to compare galaxies whose morphological features suggest they are interacting with other galaxies (mergers) to galaxies that do not seem to have been gravitationally disturbed (non-mergers). The detailed results of these statistical tests are provided in Tables 2 and 3. Here we provide a qualitative summary of our findings and we speculate about their meaning. 
\begin{itemize}
\item Galaxies that can be identified as AGN on the basis of their MIR properties are statistically different than those that cannot be identified as AGN. In particular, galaxies that host AGN have higher H$_2$/ 11.3 $\mu m$ PAH ratios and higher H$_2$ temperature as estimated from the H$_2$ S(3) /S(1) emission-line ratios.

\item The MIR emitting dust also appears affected by the presence of an AGN. We find statistically significant differences between the 11.3 $\mu$m PAH /7.7 $\mu$m PAH feature ratios of AGN and non-AGN hosts.

\item We find statistically significant differences between grain size distributions, as estimated by the 6.2$\mu$m to 7.7$\mu$m PAH feature ratios between AGN and non-AGN hosts.

\item We find statistically significant differences between the distributions of silicate strengths in AGN and non-AGN hosts: non-AGN hosts appear to have deeper absorption than AGN-hosts. 

\item We find some statistical suggestive differences between the H$_2$/ 11.3 $\mu m$ PAH ratios and H$_2$ temperature estimated from the H$_2$ S(3) /S(1) emission-line ratios of mergers and non-mergers. However, those differences are less statistically significant than those between AGN and non-AGN hosts.
\end{itemize}
Our results likely reflect the heterogenous nature of our sample and we may miss the most obscured AGN because in the nearby universe (z$<1$) the fraction of highly obscured AGN is low \citep{Lacy2007}. 
Our results do not establish a relation between the rate of BH growth and the warm ISM. However, we do find highly statistically significant differences between AGN hosts and non-AGN hosts. We do not find differences with the same high statistical significance between mergers and non-mergers.

AGN affect the ISM we measure in the IRS spectra more than gravitational interactions. This is puzzling because there is significant evidence in the literature that mergers trigger AGN and enhance star-formation \citep[e.g.][]{ellis2019}. Yet, while we do find statistical differences between the warm H$_2$ temperatures of mergers and non-mergers, those differences are less significant when we compare AGN merger hosts and AGN non-merger hosts as well as AGN/SF composites. 

One speculative explanation is that AGN in mergers were triggered by the merger, and AGNs in non-merger were triggered by secular processes. This would simply mean that the mechanisms that fuel the AGN do not fundamentally alter the ISM on kpc scales. Instead, if an AGN is present it will warm up the MIR-emitting ISM in the central kpc irrelevant of its triggering mechanism, or merger-history.  

One alternative possibility to AGN impacting the host galaxy, is that the shock heating we see in the AGN hosts' warm ISM occurs on the same time-scale as the AGN but that it is not directly linked to the AGN itself. It may be possible that the mechanisms which move gas from kiloparsec to subparsec scales (e.g supernovae, star-formation, galactic bars) also heat up the ISM. Such processes do not lead to the fast shocks measured in some mergers \citep[e.g]{app2006} but slower shocks and turbulence can enhance the temperature of the warm-molecular gas and the ratio of H$_2$ emission relative to other coolants (e.g. 11.3 $\mu$m PAH emission). We note also that AGN timescales are on the order of $10^5$ years \citep[e.g.][]{sch2015} while our ability to identify mergers is on the order of $10^6$\citep[e.g.][]{ellis2019}, so if the time-scales for shocked H$_2$ gas are similar to those of AGN then the statistics of shocked H$_2$ gas should change with AGN triggering mechanism because different AGN triggering could lead to different gas motions and properties. 

\citet{rrz2020} use SDSS optical spectra for a subsample of the \citet{lamb2019} galaxies to suggest that excess of MIR H$_2$ emission seen in AGN is produced by shocks due to AGN driven outflows in the same clouds that produce [OI] emission. Spatially resolved kinematics would be required to confirm this hypothesis. Spatially resolved studies of AGN environments in X-rays, radio, and molecular gas such as those performed by \cite{lanz2015, ogle2014, app2018} with planned and future IR facilities will help us push beyond phenomenological findings to a better physical understanding of the processes that connect AGN and ISM.

 .

{\it{Acknowledgments:}} Rebecca Minsley acknowledges support from Research Experience for Undergraduate program at the Institute for Astronomy, University of Hawaii-Manoa funded through NSF grant 6104374 and would like to thank the Institute for Astronomy for their kind hospitality during the course of this project. This research has made use of the NASA/ IPAC Infrared Science Archive, which is operated by the Jet Propulsion Laboratory, California Institute of Technology, under contract with the National Aeronautics and Space Administration.

\bibliographystyle{mnras}
\bibliography{RMrefs}

\begin{thebibliography}{}
\makeatletter
\relax
\def\mn@urlcharsother{\let\do\@makeother \do\$\do\&\do\#\do\^\do\_\do\%\do\~}
\def\mn@doi{\begingroup\mn@urlcharsother \@ifnextchar [ {\mn@doi@}
  {\mn@doi@[]}}
\def\mn@doi@[#1]#2{\def\@tempa{#1}\ifx\@tempa\@empty \href
  {http://dx.doi.org/#2} {doi:#2}\else \href {http://dx.doi.org/#2} {#1}\fi
  \endgroup}
\def\mn@eprint#1#2{\mn@eprint@#1:#2::\@nil}
\def\mn@eprint@arXiv#1{\href {http://arxiv.org/abs/#1} {{\tt arXiv:#1}}}
\def\mn@eprint@dblp#1{\href {http://dblp.uni-trier.de/rec/bibtex/#1.xml}
  {dblp:#1}}
\def\mn@eprint@#1:#2:#3:#4\@nil{\def\@tempa {#1}\def\@tempb {#2}\def\@tempc
  {#3}\ifx \@tempc \@empty \let \@tempc \@tempb \let \@tempb \@tempa \fi \ifx
  \@tempb \@empty \def\@tempb {arXiv}\fi \@ifundefined
  {mn@eprint@\@tempb}{\@tempb:\@tempc}{\expandafter \expandafter \csname
  mn@eprint@\@tempb\endcsname \expandafter{\@tempc}}}

\bibitem[\protect\citeauthoryear{{Allamandola}, {Hudgins}  \&
  {Sandford}}{{Allamandola} et~al.}{1999}]{alla1999}
{Allamandola} L.~J.,  {Hudgins} D.~M.,   {Sandford} S.~A.,  1999, \mn@doi
  [\apjl] {10.1086/311843}, \href
  {https://ui.adsabs.harvard.edu/abs/1999ApJ...511L.115A} {511, L115}

\bibitem[\protect\citeauthoryear{{Appleton} et~al.,}{{Appleton}
  et~al.}{2006}]{app2006}
{Appleton} P.~N.,  et~al., 2006, \mn@doi [\apjl] {10.1086/502646}, \href
  {http://adsabs.harvard.edu/abs/2006ApJ...639L..51A} {639, L51}

\bibitem[\protect\citeauthoryear{{Appleton} et~al.,}{{Appleton}
  et~al.}{2018}]{app2018}
{Appleton} P.~N.,  et~al., 2018, \mn@doi [\apj] {10.3847/1538-4357/aaed2a},
  \href {https://ui.adsabs.harvard.edu/abs/2018ApJ...869...61A} {869, 61}

\bibitem[\protect\citeauthoryear{{Armus} et~al.,}{{Armus}
  et~al.}{2007}]{armus2007}
{Armus} L.,  et~al., 2007, \mn@doi [\apj] {10.1086/510107}, \href
  {http://adsabs.harvard.edu/abs/2007ApJ...656..148A} {656, 148}

\bibitem[\protect\citeauthoryear{{Bally} \& {Lane}}{{Bally} \&
  {Lane}}{1982}]{bal1982}
{Bally} J.,  {Lane} A.~P.,  1982, \mn@doi [\apj] {10.1086/160017}, \href
  {http://adsabs.harvard.edu/abs/1982ApJ...257..612B} {257, 612}

\bibitem[\protect\citeauthoryear{{Becker}, {White}  \& {Helfand}}{{Becker}
  et~al.}{1994}]{vfirst}
{Becker} R.~H.,  {White} R.~L.,   {Helfand} D.~J.,  1994, in {Crabtree} D.~R.,
  {Hanisch} R.~J.,   {Barnes} J.,  eds,  Astronomical Society of the Pacific
  Conference Series Vol. 61, Astronomical Data Analysis Software and Systems
  III. p.~165

\bibitem[\protect\citeauthoryear{{Blumenthal} et~al.,}{{Blumenthal}
  et~al.}{2020}]{bb2020}
{Blumenthal} K.~A.,  et~al., 2020, \mn@doi [\mnras] {10.1093/mnras/stz3472},
  \href {https://ui.adsabs.harvard.edu/abs/2020MNRAS.492.2075B} {492, 2075}

\bibitem[\protect\citeauthoryear{{Bridge}, {Carlberg}  \& {Sullivan}}{{Bridge}
  et~al.}{2010}]{bridge2010}
{Bridge} C.~R.,  {Carlberg} R.~G.,   {Sullivan} M.,  2010, \mn@doi [\apj]
  {10.1088/0004-637X/709/2/1067}, \href
  {https://ui.adsabs.harvard.edu/abs/2010ApJ...709.1067B} {709, 1067}

\bibitem[\protect\citeauthoryear{{Chiaberge}, {Gilli}, {Lotz}  \&
  {Norman}}{{Chiaberge} et~al.}{2015}]{chia2015}
{Chiaberge} M.,  {Gilli} R.,  {Lotz} J.~M.,   {Norman} C.,  2015, \mn@doi
  [\apj] {10.1088/0004-637X/806/2/147}, \href
  {http://adsabs.harvard.edu/abs/2015ApJ...806..147C} {806, 147}

\bibitem[\protect\citeauthoryear{{Diamond-Stanic} \& {Rieke}}{{Diamond-Stanic}
  \& {Rieke}}{2012}]{ds2012}
{Diamond-Stanic} A.~M.,  {Rieke} G.~H.,  2012, \mn@doi [\apj]
  {10.1088/0004-637X/746/2/168}, \href
  {https://ui.adsabs.harvard.edu/abs/2012ApJ...746..168D} {746, 168}

\bibitem[\protect\citeauthoryear{{D{\'{\i}}az-Santos}
  et~al.,}{{D{\'{\i}}az-Santos} et~al.}{2010}]{ds2010}
{D{\'{\i}}az-Santos} T.,  et~al., 2010, \mn@doi [\apj]
  {10.1088/0004-637X/723/2/993}, \href
  {http://adsabs.harvard.edu/abs/2010ApJ...723..993D} {723, 993}

\bibitem[\protect\citeauthoryear{{Draine} \& {Li}}{{Draine} \&
  {Li}}{2007}]{dli2007}
{Draine} B.~T.,  {Li} A.,  2007, \mn@doi [\apj] {10.1086/511055}, \href
  {http://adsabs.harvard.edu/abs/2007ApJ...657..810D} {657, 810}

\bibitem[\protect\citeauthoryear{{Ellison}, {Viswanathan}, {Patton},
  {Bottrell}, {McConnachie}, {Gwyn}  \& {Cuillandre}}{{Ellison}
  et~al.}{2019}]{ellis2019}
{Ellison} S.~L.,  {Viswanathan} A.,  {Patton} D.~R.,  {Bottrell} C.,
  {McConnachie} A.~W.,  {Gwyn} S.,   {Cuillandre} J.-C.,  2019, \mn@doi
  [\mnras] {10.1093/mnras/stz1431}, \href
  {https://ui.adsabs.harvard.edu/abs/2019MNRAS.487.2491E} {487, 2491}

\bibitem[\protect\citeauthoryear{{Feigelson} \& {Nelson}}{{Feigelson} \&
  {Nelson}}{1985}]{fin1985}
{Feigelson} E.~D.,  {Nelson} P.~I.,  1985, \mn@doi [\apj] {10.1086/163225},
  \href {https://ui.adsabs.harvard.edu/abs/1985ApJ...293..192F} {293, 192}

\bibitem[\protect\citeauthoryear{{Gautier}, {Fink}, {Larson}  \&
  {Treffers}}{{Gautier} et~al.}{1976}]{gaut1976}
{Gautier} III T.~N.,  {Fink} U.,  {Larson} H.~P.,   {Treffers} R.~R.,  1976,
  \mn@doi [\apjl] {10.1086/182195}, \href
  {http://adsabs.harvard.edu/abs/1976ApJ...207L.129G} {207, L129}

\bibitem[\protect\citeauthoryear{{Genzel} et~al.,}{{Genzel}
  et~al.}{1998}]{genz1998}
{Genzel} R.,  et~al., 1998, \mn@doi [\apj] {10.1086/305576}, \href
  {https://ui.adsabs.harvard.edu/abs/1998ApJ...498..579G} {498, 579}

\bibitem[\protect\citeauthoryear{{Guillard}, {Boulanger}, {Pineau Des
  For{\^e}ts}  \& {Appleton}}{{Guillard} et~al.}{2009}]{gui2009}
{Guillard} P.,  {Boulanger} F.,  {Pineau Des For{\^e}ts} G.,   {Appleton}
  P.~N.,  2009, \mn@doi [\aap] {10.1051/0004-6361/200811263}, \href
  {http://adsabs.harvard.edu/abs/2009A%26A...502..515G} {502, 515}

\bibitem[\protect\citeauthoryear{{Guillard} et~al.,}{{Guillard}
  et~al.}{2012}]{gui2012}
{Guillard} P.,  et~al., 2012, \mn@doi [\apj] {10.1088/0004-637X/749/2/158},
  \href {http://adsabs.harvard.edu/abs/2012ApJ...749..158G} {749, 158}

\bibitem[\protect\citeauthoryear{{Hill} \& {Zakamska}}{{Hill} \&
  {Zakamska}}{2014}]{hill2014}
{Hill} M.~J.,  {Zakamska} N.~L.,  2014, \mn@doi [\mnras]
  {10.1093/mnras/stu123}, \href
  {http://adsabs.harvard.edu/abs/2014MNRAS.439.2701H} {439, 2701}

\bibitem[\protect\citeauthoryear{{Hopkins}, {Richards}  \&
  {Hernquist}}{{Hopkins} et~al.}{2007}]{hopkins2007}
{Hopkins} P.~F.,  {Richards} G.~T.,   {Hernquist} L.,  2007, \mn@doi [\apj]
  {10.1086/509629}, \href {http://adsabs.harvard.edu/abs/2007ApJ...654..731H}
  {654, 731}

\bibitem[\protect\citeauthoryear{{Hopkins}, {Torrey}, {Faucher-Gigu{\`e}re},
  {Quataert}  \& {Murray}}{{Hopkins} et~al.}{2016}]{hopkins2016}
{Hopkins} P.~F.,  {Torrey} P.,  {Faucher-Gigu{\`e}re} C.-A.,  {Quataert} E.,
  {Murray} N.,  2016, \mn@doi [\mnras] {10.1093/mnras/stw289}, \href
  {http://adsabs.harvard.edu/abs/2016MNRAS.458..816H} {458, 816}

\bibitem[\protect\citeauthoryear{{Houck} et~al.,}{{Houck}
  et~al.}{2004}]{houck2004}
{Houck} J.~R.,  et~al., 2004, in {Mather} J.~C.,  ed.,  Society of
  Photo-Optical Instrumentation Engineers (SPIE) Conference Series Vol. 5487,
  \procspie. pp 62--76, \mn@doi{10.1117/12.550517}

\bibitem[\protect\citeauthoryear{{Isobe}, {Feigelson}  \& {Nelson}}{{Isobe}
  et~al.}{1986}]{if1986}
{Isobe} T.,  {Feigelson} E.~D.,   {Nelson} P.~I.,  1986, \mn@doi [\apj]
  {10.1086/164359}, \href
  {https://ui.adsabs.harvard.edu/abs/1986ApJ...306..490I} {306, 490}

\bibitem[\protect\citeauthoryear{{Ivezi{\'c}} et~al.,}{{Ivezi{\'c}}
  et~al.}{2002}]{ivezic2002}
{Ivezi{\'c}} {\v{Z}}.,  et~al., 2002, \mn@doi [\aj] {10.1086/344069}, \href
  {https://ui.adsabs.harvard.edu/abs/2002AJ....124.2364I} {124, 2364}

\bibitem[\protect\citeauthoryear{{Kaplan} \& {Meier}}{{Kaplan} \&
  {Meier}}{1958}]{km58}
{Kaplan} E.,  {Meier} P.,  1958, J. Am. Statsitical Associations, 53, 457

\bibitem[\protect\citeauthoryear{{Lacy}, {Hill}, {Kaiser}  \&
  {Rawlings}}{{Lacy} et~al.}{1993}]{lacy1993}
{Lacy} M.,  {Hill} G.~J.,  {Kaiser} M.-E.,   {Rawlings} S.,  1993, \mn@doi
  [\mnras] {10.1093/mnras/263.3.707}, \href
  {https://ui.adsabs.harvard.edu/abs/1993MNRAS.263..707L} {263, 707}

\bibitem[\protect\citeauthoryear{{Lacy}, {Sajina}, {Petric}, {Seymour},
  {Canalizo}, {Ridgway}, {Armus}  \& {Storrie-Lombardi}}{{Lacy}
  et~al.}{2007}]{Lacy2007}
{Lacy} M.,  {Sajina} A.,  {Petric} A.~O.,  {Seymour} N.,  {Canalizo} G.,
  {Ridgway} S.~E.,  {Armus} L.,   {Storrie-Lombardi} L.~J.,  2007, \mn@doi
  [\apjl] {10.1086/523851}, \href
  {http://adsabs.harvard.edu/abs/2007ApJ...669L..61L} {669, L61}

\bibitem[\protect\citeauthoryear{{Lambrides}, {Petric}, {Tchernyshyov},
  {Zakamska}  \& {Watts}}{{Lambrides} et~al.}{2019}]{lamb2019}
{Lambrides} E.~L.,  {Petric} A.~O.,  {Tchernyshyov} K.,  {Zakamska} N.~L.,
  {Watts} D.~J.,  2019, \mn@doi [\mnras] {10.1093/mnras/stz1316}, \href
  {https://ui.adsabs.harvard.edu/abs/2019MNRAS.tmp.1261L} {p.~1261}

\bibitem[\protect\citeauthoryear{{Lanz}, {Ogle}, {Evans}, {Appleton},
  {Guillard}  \& {Emonts}}{{Lanz} et~al.}{2015}]{lanz2015}
{Lanz} L.,  {Ogle} P.~M.,  {Evans} D.,  {Appleton} P.~N.,  {Guillard} P.,
  {Emonts} B.,  2015, \mn@doi [\apj] {10.1088/0004-637X/801/1/17}, \href
  {https://ui.adsabs.harvard.edu/abs/2015ApJ...801...17L} {801, 17}

\bibitem[\protect\citeauthoryear{{Larkin}, {Armus}, {Knop}, {Soifer}  \&
  {Matthews}}{{Larkin} et~al.}{1998}]{lark1998}
{Larkin} J.~E.,  {Armus} L.,  {Knop} R.~A.,  {Soifer} B.~T.,   {Matthews} K.,
  1998, \mn@doi [\apjs] {10.1086/313063}, \href
  {http://adsabs.harvard.edu/abs/1998ApJS..114...59L} {114, 59}

\bibitem[\protect\citeauthoryear{{Larson} et~al.,}{{Larson}
  et~al.}{2016}]{larson2016}
{Larson} K.~L.,  et~al., 2016, \mn@doi [\apj] {10.3847/0004-637X/825/2/128},
  \href {https://ui.adsabs.harvard.edu/abs/2016ApJ...825..128L} {825, 128}

\bibitem[\protect\citeauthoryear{Lebouteiller, Barry, Spoon, Bernard-Salas,
  Sloan, Houck  \& Weedman}{Lebouteiller et~al.}{2011}]{cas2011}
Lebouteiller V.,  Barry D.~J.,  Spoon H. W.~W.,  Bernard-Salas J.,  Sloan
  G.~C.,  Houck J.~R.,   Weedman D.~W.,  2011, \mn@doi [The Astrophysical
  Journal Supplement Series] {10.1088/0067-0049/196/1/8}, 196, 8

\bibitem[\protect\citeauthoryear{{Li} \& {Draine}}{{Li} \&
  {Draine}}{2001}]{lid2001}
{Li} A.,  {Draine} B.~T.,  2001, \mn@doi [\apj] {10.1086/323147}, \href
  {http://adsabs.harvard.edu/abs/2001ApJ...554..778L} {554, 778}

\bibitem[\protect\citeauthoryear{{Moorwood} \& {Oliva}}{{Moorwood} \&
  {Oliva}}{1988}]{mor1988}
{Moorwood} A.~F.~M.,  {Oliva} E.,  1988, \aap, \href
  {http://adsabs.harvard.edu/abs/1988A%26A...203..278M} {203, 278}

\bibitem[\protect\citeauthoryear{{Nair} \& {Abraham}}{{Nair} \&
  {Abraham}}{2010}]{nair&abr2010}
{Nair} P.~B.,  {Abraham} R.~G.,  2010, \mn@doi [\apjs]
  {10.1088/0067-0049/186/2/427}, \href
  {https://ui.adsabs.harvard.edu/abs/2010ApJS..186..427N} {186, 427}

\bibitem[\protect\citeauthoryear{{Nesvadba}, {Boulanger}, {Lehnert}, {Guillard}
   \& {Salome}}{{Nesvadba} et~al.}{2011}]{nesvadba2011}
{Nesvadba} N.~P.~H.,  {Boulanger} F.,  {Lehnert} M.~D.,  {Guillard} P.,
  {Salome} P.,  2011, \mn@doi [\aap] {10.1051/0004-6361/201118018}, \href
  {https://ui.adsabs.harvard.edu/abs/2011A&A...536L...5N} {536, L5}

\bibitem[\protect\citeauthoryear{{Ogle}, {Antonucci}, {Appleton}  \&
  {Whysong}}{{Ogle} et~al.}{2007}]{ogle2007}
{Ogle} P.,  {Antonucci} R.,  {Appleton} P.~N.,   {Whysong} D.,  2007, \mn@doi
  [\apj] {10.1086/521334}, \href
  {http://adsabs.harvard.edu/abs/2007ApJ...668..699O} {668, 699}

\bibitem[\protect\citeauthoryear{{Ogle}, {Lanz}  \& {Appleton}}{{Ogle}
  et~al.}{2014}]{ogle2014}
{Ogle} P.~M.,  {Lanz} L.,   {Appleton} P.~N.,  2014, \mn@doi [\apjl]
  {10.1088/2041-8205/788/2/L33}, \href
  {https://ui.adsabs.harvard.edu/abs/2014ApJ...788L..33O} {788, L33}

\bibitem[\protect\citeauthoryear{{Petric} et~al.,}{{Petric}
  et~al.}{2011}]{petric2011}
{Petric} A.~O.,  et~al., 2011, \mn@doi [\apj] {10.1088/0004-637X/730/1/28},
  \href {http://adsabs.harvard.edu/abs/2011ApJ...730...28P} {730, 28}

\bibitem[\protect\citeauthoryear{Petric et~al.,}{Petric
  et~al.}{2018}]{petric2018}
Petric A.~O.,  et~al., 2018, The Astronomical Journal, 156, 295

\bibitem[\protect\citeauthoryear{{Riffel}, {Zakamska}  \& {Riffel}}{{Riffel}
  et~al.}{2020}]{rrz2020}
{Riffel} R.~A.,  {Zakamska} N.~L.,   {Riffel} R.,  2020, \mn@doi [\mnras]
  {10.1093/mnras/stz3137}, \href
  {https://ui.adsabs.harvard.edu/abs/2020MNRAS.491.1518R} {491, 1518}

\bibitem[\protect\citeauthoryear{{Rigopoulou}, {Kunze}, {Lutz}, {Genzel}  \&
  {Moorwood}}{{Rigopoulou} et~al.}{2002}]{rig2002}
{Rigopoulou} D.,  {Kunze} D.,  {Lutz} D.,  {Genzel} R.,   {Moorwood} A.~F.~M.,
  2002, \mn@doi [\aap] {10.1051/0004-6361:20020607}, \href
  {https://ui.adsabs.harvard.edu/abs/2002A&A...389..374R} {389, 374}

\bibitem[\protect\citeauthoryear{{Roussel} et~al.,}{{Roussel}
  et~al.}{2007}]{rous2007}
{Roussel} H.,  et~al., 2007, \mn@doi [\apj] {10.1086/521667}, \href
  {http://adsabs.harvard.edu/abs/2007ApJ...669..959R} {669, 959}

\bibitem[\protect\citeauthoryear{{Schawinski}, {Koss}, {Berney}  \&
  {Sartori}}{{Schawinski} et~al.}{2015}]{sch2015}
{Schawinski} K.,  {Koss} M.,  {Berney} S.,   {Sartori} L.~F.,  2015, \mn@doi
  [\mnras] {10.1093/mnras/stv1136}, \href
  {https://ui.adsabs.harvard.edu/abs/2015MNRAS.451.2517S} {451, 2517}

\bibitem[\protect\citeauthoryear{{Smith} et~al.,}{{Smith}
  et~al.}{2007}]{smith07}
{Smith} J.~D.~T.,  et~al., 2007, \mn@doi [\pasp] {10.1086/522634}, \href
  {http://adsabs.harvard.edu/abs/2007PASP..119.1133S} {119, 1133}

\bibitem[\protect\citeauthoryear{{Stierwalt} et~al.,}{{Stierwalt}
  et~al.}{2013}]{stir2013}
{Stierwalt} S.,  et~al., 2013, \mn@doi [\apjs] {10.1088/0067-0049/206/1/1},
  \href {http://adsabs.harvard.edu/abs/2013ApJS..206....1S} {206, 1}

\bibitem[\protect\citeauthoryear{{Stierwalt} et~al.,}{{Stierwalt}
  et~al.}{2014}]{stir2014}
{Stierwalt} S.,  et~al., 2014, \mn@doi [\apj] {10.1088/0004-637X/790/2/124},
  \href {http://adsabs.harvard.edu/abs/2014ApJ...790..124S} {790, 124}

\bibitem[\protect\citeauthoryear{{Werner} et~al.,}{{Werner}
  et~al.}{2004}]{werner04}
{Werner} M.~W.,  et~al., 2004, \mn@doi [\apjs] {10.1086/422992}, \href
  {http://adsabs.harvard.edu/abs/2004ApJS..154....1W} {154, 1}

\bibitem[\protect\citeauthoryear{{Zakamska}}{{Zakamska}}{2010}]{zakamska2010}
{Zakamska} N.~L.,  2010, \mn@doi [\nat] {10.1038/nature09037}, \href
  {https://ui.adsabs.harvard.edu/abs/2010Natur.465...60Z} {465, 60}

\bibitem[\protect\citeauthoryear{{Zheng}, {Kriss}, {Telfer}, {Grimes}  \&
  {Davidsen}}{{Zheng} et~al.}{1997}]{zheng1997}
{Zheng} W.,  {Kriss} G.~A.,  {Telfer} R.~C.,  {Grimes} J.~P.,   {Davidsen}
  A.~F.,  1997, \apj, \href {http://adsabs.harvard.edu/abs/1997ApJ...475..469Z}
  {475, 469}

\makeatother
\end{thebibliography}
\end{document}